\documentclass[12pt]{article}

\usepackage{amsfonts}
\usepackage{amsmath}
\usepackage{graphicx}
\usepackage[english]{babel}
\usepackage[]{graphicx}
\usepackage{color}
\usepackage{hyperref}
\usepackage{bbm}

\title{Spherical Symmetric Solutions in Ho\v{r}ava-Lifshitz Gravity and their Properties}
\author{D.~Capasso\footnote{dcapass00@ccny.cuny.edu}\\  
Physics Department, City College of the CUNY\\
160 Convent Avenue, New York, NY 10031}

\begin{document}

\maketitle
\begin{abstract}
Non-projectable Ho\v{r}ava gravity for a spherically symmetric configuration with $\lambda=1$ exhibits an infinite set of solutions parametrized by a generic function $g^{2}(r)$ for the radial component of the shift vector. In the IR limit the infinite set of solutions corresponds to the invariance of General Relativity under a spacetime reparametrization. In general, not being a coordinate transformation, the symmetry in the action responsible for the infinite set of solutions does not have a clear physical interpretation. Indeed it is broken by the matter term in the action. We study the behavior of the solutions for generic values of the parameter $g^{2}(r)$.
\end{abstract}

\tableofcontents

\section{Introduction}\label{sec:intro}
Ho\v{r}ava-Lifshitz (HL) gravity \cite{Horava:MQC,Horava:QGLP} is a non-relativistic extension of General Relativity (GR) which aims to recover GR in the infrared (IR) limit. Its main characteristic is its invariance under the anisotropic
rescaling
\[
x\to bx \qquad t\to b^{z}t,
\]
which makes the conformal dimension ($[\phantom{-}]_{s}$) of space and time to be different:
\[
[x]_{s}=-1 \qquad [t]_{s}=-z.
\]
This is to achieve power-counting renormalizability, making HL gravity a UV completion of Hilbert-Einstein Gravity. For a $(3+1)$-dimensional space-time $z=3$ \cite{Horava:QGLP}.

To implement the anisotropic scaling the theory is constructed on a space-time of the form $M=\mathbb{R}\times\Sigma$, where $\Sigma$ is a space-like $3$-dimensional surface. The action is given by the usual Hilbert-Einstein action, written in ADM components, plus  higher powers or higher spatial derivatives of the intrinsic curvature of $\Sigma$  to modify the potential term\footnote{The action considered here is a modified one, introduced in \cite{KehagiasSfetsos:BHFRWGNRG}.}:
\begin{eqnarray}\label{KSaction}
S &=&\int dtd^{3}x\sqrt{h}N\left\{\frac{2}{\kappa^{2}}(K_{ij}K^{ij}-\lambda K^{2})
+\mu^{4}\mathcal{R}
-\frac{\kappa^{2}}{2w^{4}}C_{ij}C^{ij}
+\frac{\kappa^{2}\mu}{2w^{2}}\epsilon^{ijk}\mathcal{R}_{il}\nabla_{j}\mathcal{R}^{l}_{\phantom{-}k}+\right.\nonumber\\
&& \left.-\frac{\kappa^{2}\mu^{2}}{8}\mathcal{R}_{ij}\mathcal{R}^{ij}
+\frac{\kappa^{2}\mu^{2}}{8(1-3\lambda)}\left(\frac{1-4\lambda}{4}\mathcal{R}^{2}+\Lambda_{W}\mathcal{R}-3\Lambda_{W}^{2}\right)
\right\}
\end{eqnarray}
where the kinetic term corresponds to the first bracket, in which
\[
K_{ij}=\frac{1}{2N}(\dot{h}_{ij}-\nabla_{i}N_{j}-\nabla_{j}N_{i})
\]
is the extrinsic curvature, in which $N$ is the lapse function, $N_{i}$ the shift vector and $h_{ij}$ the spatial metric on $\Sigma$; in the potential term $\mathcal{R}_{ij}$ is the Ricci intrinsic curvature on $\Sigma$, $\mathcal{R}$ is its trace and $C^{ij}=\varepsilon^{ikl}\nabla_{k}\left(R_{l}^{\phantom{l}j}-\frac{R}{4}\delta_{l}^{\phantom{l}j}\right)$ is the Cotton tensor. The IR limit is given by retaining the kinetic term with $\lambda=1$, the term $\mu^{4}\mathcal{R}$ and, eventually, the constant term that will reproduce the cosmological constant. Quantizing HL gravity the coupling constants will be running, therefore it is supposed that in the IR limit $w\to\infty$, $\lambda$ goes to $1$ and that the higher powers of the spatial curvature are negligible. This reproduces the Hilbert-Einstein action with $z=1$, thus recovering the relativistic isotropy of spacetime.

The potential term was first introduced using the detailed balance condition \cite{Horava:QGLP}; more general expressions were considered in \cite{KiritsisKofinas:HLC,SotiriouVisserWeinfurtner:PLVQG,SotiriouVisserWeinfurtner:QGWLI}. In particular, Kehagias and Sfetsos considered in \cite{KehagiasSfetsos:BHFRWGNRG} an action obtained by softly breaking the detailed balance condition with a
curvature term $\mu^{4}\mathcal{R}$. Here we will consider the expression (relation~(5) in \cite{CapassoPolychronakos:GSSSSHG} with a zero cosmological constant)
\begin{eqnarray}
S &=& \frac{\kappa^2 \mu^2}{8} \int dtd^{3}x\sqrt{h}N\Bigl\{
\omega (K_{ij}K^{ij}-\lambda K^{2})
+\omega\mathcal{R}
- \frac{4}{\mu^2 w^4} C_{ij}C^{ij}+
\nonumber\\
&&+\frac{4}{\mu w^2}
\sqrt{3\lambda-1}\epsilon^{ijk}\mathcal{R}_{il}\nabla_{j}\mathcal{R}^{l}_{\phantom{-}k}
 -(3\lambda -1) \mathcal{R}_{ij}\mathcal{R}^{ij}
+\frac{4\lambda-1}{4}\mathcal{R}^{2}
\Bigr\}
\end{eqnarray}
obtained from the KS action (\ref{KSaction}) with some redefinitions of the coupling constants (see \cite{CapassoPolychronakos:GSSSSHG}).

In the literature two versions of HL gravity are usually considered: the non-projectable case, in which the lapse function has a full space-time dependence, and the projectable case, in which the lapse function depends only on the time coordinate. The non-projectable case, which is the one considered in this article, suffers from a strong coupling problem \cite{CharmousisNizPadillaSaffin:SCHG,BlasPujolasSibiryakov:EMIHG} which might make the theory unstable in the present formulation; moreover the unstable scalar mode does not decouple in the IR limit, therefore GR is not really recovered. The projectable case, on the other hand, is more tractable although it still suffers from a strong coupling problem \cite{CharmousisNizPadillaSaffin:SCHG}. Here we will not consider the non-projectable theory because it is overly constrained and does not reproduce the Schwarzschild spherically symmetric solution.

Several aspects of the Kehagias-Sfetsos action were analyzed in the literature: cosmological solutions \cite{Park:BHCSIRMHG,Park:BHSIRMHG,CaiSaridakis:NSCNRG,AliDuttaSaridakisSen:HLCCG}, possible tests \cite{Konoplya:TCHLG,HarkoKivacsLobo:SSTHLG,Park:THGDE,IorioRuggiero:HLGSSOM,IorioRuggiero:CKSHLG}, fundamental aspects of the theory \cite{LiPang:THLG,Kobakhidze:IRHGHC,ChenHuang:FTLP,OrlandoReffert:RHLG,Suyama:2009vy,CapassoPolychronakos,Sindoni:PKHL,Rama:PMHLDR,Suyama:NMHLG,RomeroCuestaGarciaVergara:CAM}, black hole solutions (with vanishing shift variables) \cite{LuMeiPope:SHG,MyungKim:THLBH,CaiCaoOhta:TBHHLG,CaiCaoOhta:ThBHHLG,GhodsiHatefi:ERSHG,Myung:TBHDHLG,Myung:EBHDHLG,
LeeKimMyung:EBHHLG,PengWu:HRBHIMHLG,KiritsisKofinas:HLBH,TangChen:SSSSMHLGPC,CaiLiuSun:z=4HLG}, special cases such as $\lambda=1/3$ \cite{Park:HGCP} and possible extensions \cite{KoutsoumbasPapantonopoulosPasipoularidesTsoukalas:BHS5DHLG,KoutsoumbasPasipoularides:BHSHLGC}. In particular Kiritsis and Kofinas in \cite{KiritsisKofinas:HLBH} studied more general solutions considering the Ho\v{r}ava-Lifshitz action with generic (independent) coupling constants, that is, an action not derived from a detailed balance condition.

In the present work we are interested in studying the infinite set of spherically symmetric solutions for the case $\lambda=1$ (reviewed in sec.~\ref{sec:lambda=1}). The choice $\lambda=1$ is dictated not only by the fact that it is the expected value in the IR limit, but also by the fact that the gauge invariance found in \cite{CapassoPolychronakos:GSSSSHG} could be used to fix $\lambda=1$ from the beginning once the theory is quantized (there are also different works in which it is shown how, leaving $\lambda$ general, it is possible to achieve a correct IR limit introducing second order constraints \cite{BellorinRestuccia:CHT}). A promising advance in this direction, although it is not clear yet if there is any relation with our gauge symmetry, was made by Ho\v{r}ava and Melby-Thompson in \cite{HoravaThompson:GCQGLP}. In \cite{AlexandrePasipoularides:SSSCHLG} spherically symmetric solutions are discussed.

Our gauge symmetry is manifestly broken by the interaction with matter\footnote{Here we are considering the usual relativistic matter term, although in principle it may be possible to consider a deformed term that is gauge invariant (for more general interactions terms see \cite{CapassoPolychronakos,Sindoni:PKHL,Rama:PMHLDR,Suyama:NMHLG,RomeroCuestaGarciaVergara:CAM}).} therefore for each value of the gauge-parameter $g(r)$ we have physically different solutions. In section~\ref{sec:constraints_g2} we will study the constraints to which $g(r)$ is subject to have a well defined metric, while in sections~\ref{sec:bending} and \ref{sec:singularity} we will review some physical aspects of the problem, the possible measurement of $g(r)$ from astrophysical data and the behavior of spacetime behind the horizon.

\section{Spherically Symmetric Solutions for $\lambda=1$}\label{sec:lambda=1}
As shown in the appendix~\ref{sec:appendix} and in \cite{CapassoPolychronakos:GSSSSHG}, in HL gravity, unlike GR, the non-diagonal metric 
\begin{equation}\label{ansatz}
ds^{2}=
-(N^{2}-N_{r}^{2}f)dt^{2}
+2N_{r}drdt
+\frac{dr^{2}}{f}
+r^{2}d\theta^{2}+r^{2}\sin^{2}\theta d\phi^{2}
\end{equation}
and the diagonal one
\begin{equation}\label{diagonalmetric}
ds^{2}=
-N^{*2}dt^{*2}
+\frac{1}{f^{*}}dr^{*2}
+r^{*2}d\theta^{2}+r^{*2}\sin^{2}\theta d\phi^{2}
\end{equation}
are not equivalent because we cannot perform the relevant coordinate transformation:
\begin{equation}\label{nd-to-d}
\begin{array}{c}
dt=dt^{*}+\frac{N_{r}}{N^{2}-N_{r}^{2}f}dr\\ \\
\textrm{with}\qquad
N^{*2}=N^{2}-N_{r}^{2}f \qquad
f^{*}=\frac{f(N^{2}-N_{r}^{2}f)}{N^{2}}.
\end{array}
\end{equation}
Hence the most general spherically symmetric ansatz in HL gravity is (\ref{ansatz}). We call these ``hedgehog'' solutions, in analogy with the field theoretic soliton configurations of the same name, as they possess radially-pointing ``hair'' due to the shift field.

In \cite{CapassoPolychronakos:GSSSSHG} we showed that, for the ansatz (\ref{ansatz}) and for $\lambda=1$, we have an infinite set of solutions:
\begin{equation}\label{g-solution}
f=1+\omega r^{2}
-\sqrt{\omega^{2}r^{4}+4\omega Mr-2\omega g^{2}r^{2}} \qquad
N^{2}=f \qquad
N_{r}=\pm\sqrt{\frac{g^{2}}{f}}
\end{equation}
where $g^{2}$ is a generic function of the radial coordinate that parametrizes the set of solutions. Here we will assume $f$ and $g^{2}$ to be analytic functions.

It is evident that these solutions are well defined only for
\[
f>0 \qquad\textrm{and}\qquad
g^{2}\leq\frac{\omega}{2}r^{2}+\frac{2M}{r}.
\]
In the following, and in particular in section~\ref{sec:constraints_g2}, we will find other constraints on $g^{2}$ to have a well behaved metric.

Although we cannot consider the two form of the metric, (\ref{ansatz}) and (\ref{diagonalmetric}), to be physically equivalent in this context, we can still perform the coordinate transformation (\ref{nd-to-d}) to obtain intermediate expressions and then go back to the non-diagonal coordinates to study the physical results. In all the cases we consider here the relations used are relativistic because we consider the standard relativistic coupling with matter, therefore, in the diagonal coordinates, the expressions for the equations of motion for a test particle, the bending of light or the relation for the position of the horizon are exactly like the one obtained in the GR context, only in the wrong coordinate system. This means that we can just use the relativistic relation and perform the change of coordinates to go to non-diagonal coordinates to obtain the result we seek.

In our case the  the coordinate transformation becomes
\begin{equation}\label{Ftransf}
dt=dt^{*}+\sqrt{\frac{g^{2}}{f}}\frac{1}{(f-g^{2})}dr;
\end{equation}
which is defined only for
\begin{equation}\label{Fconditions}
f>0 \qquad\textrm{and}\qquad f\neq g^{2} \quad(f^{*}\neq0).
\end{equation}

The first condition, $f>0$, implies a constraint on $g^{2}$
\begin{equation}\label{f>0}
g^{2}\geq-1+\frac{2M}{r}-\frac{1}{2\omega r^{2}},
\end{equation}
while the second condition, as already noted in \cite{CapassoPolychronakos:GSSSSHG}, is related to the position $r_{h}$ of the horizon, which is obtained by solving the condition
\begin{equation}
f^{*}(r_{h})=0 \qquad\Rightarrow\qquad f(r_{h})=g^{2}(r_{h})
\end{equation}
and choosing the outer solution. In particular, from the relations above, it is easy to show that
\begin{equation}\label{rh<=2M}
0\leq(g^{2}(r_{h})-1)^{2}
=2\omega r_{h}(2M-r_{h}),
\end{equation}
which corresponds to say that no horizon beyond the Schwarzschild radius is possible ($r_{h}\leq2M$).

Under the conditions (\ref{Fconditions}) the coefficients in the diagonal metric become:
\begin{equation}\label{f*}
N^{*2}=f^{*}
=f-g^{2}
=1+\omega r^{2}
-\sqrt{\omega^{2}r^{4}+4\omega Mr-2\omega g^{2}r^{2}}
-g^{2}.
\end{equation}

The implications of conditions (\ref{Fconditions}) are that the change of coordinates is allowed only in the region outside the horizon, exactly as in GR. With an opportune change of coordinates in GR we can extend the spacetime inside the horizon, but it is not clear how to proceed in HL gravity, where the foliation has a geometrical meaning, not the metric. We will briefly investigate this problem in section~\ref{sec:singularity}.

\section{Constraints on $g^{2}$}\label{sec:constraints_g2}

Let us forget for a moment about the problems related to the extension of the metric inside the horizon and concentrate on analyzing the infinite set of all possible solutions to the equations of motion, looking for the constraints which the generic function $g$ must obey to reproduce a well behaved metric.

The solutions in \cite{CapassoPolychronakos:GSSSSHG} for $\lambda=1$ exhibit the following invariance:
\begin{equation}\label{deltag}
\frac{\delta_{g}N^{2}}{N^{2}}=\frac{\delta_{g}f}{f} \qquad
\delta_{g}N_{r}^{2}
=\frac{1+\omega r^{2}-f-\omega r^{2}N_{r}^{2}f/N^{2}}{\omega r^{2}f}
\frac{N^{2}\delta_{g}f}{f}.
\end{equation}
This invariance is not related to a coordinate transformation (although it recovers the usual GR coordinate invariance in the limit $\omega r^{2}\to\infty$) and therefore does not admit yet a simple physical interpretation. Moreover, as will be noted in section~\ref{sec:bending}, this invariance is not a symmetry of the matter term. As a consequence we need to treat each possible value of $g$ as the source of a different solution. In this context it makes sense to discard some of the solutions based on the study of the constraints that $g^{2}$ must fulfill.

As already observe the first constraint on $g^{2}$ comes from the expression (\ref{f*}) for which we must have
\begin{equation}\label{g2constraint}
g^{2}\leq\frac{\omega}{2}r^{2}+\frac{2M}{r}.
\end{equation}
This, in particular, implies
\begin{equation}\label{g2r2}
g^{2}r^{2}|_{r=0}=0.
\end{equation}
The constraint (\ref{f>0}) is always compatible with the upper bound for $g^{2}$ since
\[
-1+\frac{2M}{r}-\frac{1}{2\omega r^{2}}\leq\frac{\omega}{2}r^{2}+\frac{2M}{r}
\]
and is relevant only when the l.h.s is positive (remember that $g^{2}>0$):
\[
-1+\frac{2M}{r}-\frac{1}{2\omega r^{2}}>0 \qquad\Rightarrow\qquad
M-\sqrt{M^{2}-\frac{1}{2\omega}}<r<M+\sqrt{M^{2}-\frac{1}{2\omega}},
\]
that is, between the internal and the external horizon radius for the KS metric. This in particular means that only when $f^{*}\leq0$ the condition $f>0$ may becomes relevant. Indeed, we have that $f$ for a generic value of $g^{2}$ satisfies
\[
1+\omega r^{2}-\sqrt{\omega^{2}r^{4}+4\omega Mr}
\leq f<
1+\omega r^{2}
\]
The limits correspond, respectively, to $g^{2}=0$ and $g^{2}=\frac{\omega}{2}r^{2}+\frac{2M}{r}$.

Also if we consider only a constrained subclass of possible $g^{2}$, it is still an infinite set. We can then study the range in which $f^{*}$ may vary for different $g^{2}$'s.

For $g=0$ we have the well known KS solution
\[
f_{0}^{*}=1+\omega r^{2}
-\sqrt{\omega^{2}r^{4}+4\omega Mr};
\]
while for $g^{2}=\frac{\omega}{2}r^{2}+\frac{2M}{r}$ we have
\[
f^{*}_{max}=1-\frac{2M}{r}+\frac{\omega}{2}r^{2},
\]
that is, a de Sitter-like solution. The inequality
\[
f^{*}_{max}\geq 1+\omega r^{2}
-\sqrt{\omega^{2}r^{4}+4\omega Mr-2\omega g^{2}r^{2}}
-g^{2}
\]
is true for
\[
-\frac{3}{2}\omega r^{2}+\frac{2M}{r}\leq g^{2}
\]
which is always verified for
\[
-\frac{3}{2}\omega r^{2}+\frac{2M}{r}\leq0,
\]
that is,
\[
r\geq\left(
\frac{4M}{3\omega}
\right)^{1/3}.
\]
Then, restricting to the case $r\gg \left(\frac{4M}{3\omega}\right)^{1/3}$, which for $\omega\gg1$ is much less than the classical horizon $2M$, we can consider $f^{*}_{max}$ to be the upper bound for $f^{*}$.

\begin{figure}[h]\label{fig:f*}
\begin{center}
\includegraphics[scale=.5]{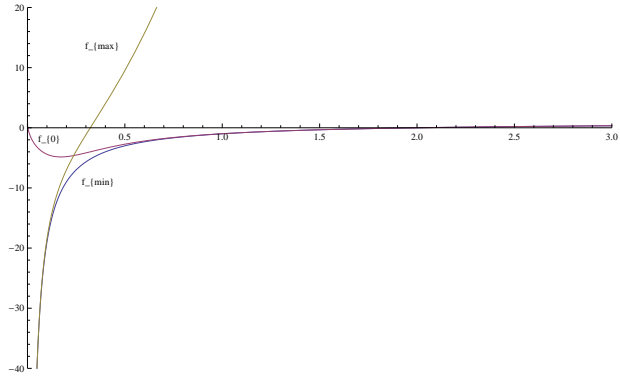}
\end{center}
\caption{In this figure we plot $f_{min}^{*}$, $f_{max}^{*}$ and $f_{0}^{*}$ with $\tilde{r}=r/M$ and $\tilde{\omega}=\omega M^{2}=100$.}
\end{figure}

Minimizing $f^{*}$ with respect to $g$ we find two values
\[
g=0 \qquad\textrm{and}\qquad g=\sqrt{\frac{2M}{r}}.
\]
For $g^{2}=\frac{2M}{r}$ we have
\begin{equation}\label{fmin}
f^{*}_{min}=1-\frac{2M}{r}.
\end{equation}
Therefore the Schwarzschild solution is the lower bound solution for the metric (fig.\ref{fig:f*}).

Finally note that redefining $\omega$ and $r$ as dimensionless quantities, $\tilde{\omega}=\omega M^{2}$ and $\tilde{r}=r/M$, we can rewrite $f^{*}$ as
\[
f^{*}=f-g^{2}
=1+\tilde{\omega}\tilde{r}^{2}
-\sqrt{\tilde{\omega}^{2}\tilde{r}^{4}+4\tilde{\omega}\tilde{r}-2\tilde{\omega}g^{2}\tilde{r}^{2}}
-g^{2}
\]
that shows that the mass $M$ is just a scaling parameter.

\subsection{Asymptotic Behavior}
In this section we want to find what are the restrictions to the asymptotic behavior of $g^{2}$ to have a flat asymptotic space-time. Let us then consider $f^{*}(r)$ for $r\to\infty$. Writing
\[
f^{*}
=1+\omega r^{2}\left[
1-\sqrt{1+\frac{4M}{\omega r^{3}}-\frac{2g^{2}}{\omega r^{2}}}
\right]
-g^{2}
\]
it is easy two see that we can encounter three cases for $r\to\infty$: the terms $\frac{2g^{2}}{\omega r^{2}}$ is negligible with respect to $\frac{4M}{\omega r^{3}}$, it is comparable with $\frac{4M}{\omega r^{3}}$ or $\frac{4M}{\omega r^{3}}$ is negligible with respect to $\frac{2g^{2}}{\omega r^{2}}$. In the first two cases we can directly obtain that the asymptotic limit is flat spacetime; in the third case we can neglect the term $\frac{4M}{\omega r^{3}}$. In this case, considering that asymptotically $2g^{2}\leq\omega r^{2}$, we can expand $f^{*}$ in terms of $g^{2}$ as follows:
\[
f^{*}
=1+\sum_{n=2}\frac{(2n-3)!!}{n!}\frac{g^{2n}}{\omega^{n-1}r^{2n-2}}.
\]
The above relation implies that $g^{2}$, asymptotically, must grow less fast than $Cr$ to have a Minkowski flat asymptotic space-time.

For $g^{2}=Cr$ the function $f^{*}$ asymptotically goes as
\[
1+\sum_{n=2}\frac{(2n-3)!!}{n!}\frac{C^{n}}{\omega^{n-1}}r^{2-n}
\simeq 1+\frac{C^{2}}{2\omega};
\]
in this case the asymptotic metric interval becomes
\begin{equation}\label{g2=Cr}
ds^{2}
=-(1+\frac{C^{2}}{2\omega})dt^{*2}+\frac{dr^{2}}{1+\frac{C^{2}}{2\omega}}+r^{2}d\Omega
\end{equation}
that, after a rescaling of the time and of the radial coordinate ($\sqrt{1+\frac{C^{2}}{2\omega}}t^{*}\to t^{*}$ and $r\to\sqrt{1+\frac{C^{2}}{2\omega}}r$), becomes
\[
ds^{2}
=-dt^{*2}+dr^{2}+(1+\frac{C^{2}}{2\omega})r^{2}d\Omega
\]
showing a conical singularity.

Such a space is asymptotically flat in the sense that there exists a coordinate transformation (\ref{Ftransf}) for which the diagonal metric is Mikowsky-like. Going back to the time coordinate determined by the foliation the metric (\ref{g2=Cr}) asymptotically becomes
\[
ds^{2}
=-(1+\frac{C^{2}}{2\omega})dt^{2}
+2drdt
+dr^{2}
+r^{2}d\Omega.
\]
To have a true flat asymptotic space-time we need to impose the condition $\lim_{r\to\infty}N_{r}=0$ that simply implies $\lim_{r\to\infty}g(r)=0$, being $\lim_{r\to\infty}f(r)$ finite by construction. Under this condition we directly obtain a Minkowskian asymptotic behavior.

\section{The Bending of Light Measure}\label{sec:bending}

It is easy to show that in both coordinate frames the Killing vectors that correspond, respectively, to the energy $E$ and to the angular momentum $L$ take the same form
\[
\boldsymbol{\xi}_{E}=\partial_{t}=\partial_{t^{*}} \qquad
\boldsymbol{\xi}_{L}=\partial_{\phi}.
\]
In particular the energy is given by
\[
E=-p_{0}^{*}
=N^{*2}p^{*0}
=(N^{2}-N_{r}^{2}f)\left[
p^{0}-\frac{N_{r}p^{r}}{N^{2}-N_{r}^{2}f}
\right]
=(N^{2}-N_{r}^{2}f)p^{0}-N_{r}p^{r}
\equiv-p_{0}.
\]

Therefore for a particle of mass $m$ the dispersion relation yields
\[
-\frac{\varepsilon^{2}}{N^{*2}}+\frac{\dot{r}^{2}}{f^{*}}+\frac{l^{2}}{r^{2}}+1=0,
\]
where
\[
\varepsilon=E/m \qquad l=L/m \qquad
k=\left\{\begin{array}{cc}
1 & \textrm{massive particle}\\
0 & \textrm{massless particle}\\
\end{array}\right..
\]
The equations of motion are
\begin{eqnarray}\label{EM}
\varepsilon &=& (N^{2}-N_{r}^{2}f)\dot{t}-N_{r}\dot{r}\\
l &=& r^{2}\dot{\phi}\\
\dot{r}^{2} &=& \varepsilon^{2}-V_{eff}
\end{eqnarray}
where we defined the effective potential as
\[
V_{eff}=f^{*}\left(
\frac{l^{2}}{r^{2}}+1
\right).
\]


The standard matter term in the action is not invariant under the gauge transformation (\ref{deltag}), so metrics with different $g$'s represent physically different solutions. This is also evident from the radial equations of motion (\ref{EM}) that depend on $g^{2}$: the trajectory depends on $g^{2}$ although the radial coordinates is not involved in the transformation (\ref{deltag}). Therefore the only way we have to fix $g^{2}$ is to study the trajectory of a test particle and reconstruct from it the function $g^{2}$: the measurement of the bending of light as a function of the radial coordinate $r$, that is $\frac{d\phi}{dr}$, and the impact parameter can be used to determine the metric.

For a ray of light the radial equation of motion is
\[
\dot{r}^{2}=E^{2}-\frac{L^{2}}{r^{2}}f^{*}
\]
and the impact parameter is defined as usual as
\[
b=\frac{L}{E}=\left.\sqrt{\frac{r^{2}}{f^{*}}}\right|_{r=R_{0}}
\]
where $R_{0}$ is the closest distance to the star in the trajectory. Using the equation of motion for $\phi$, as in GR, we obtain
\[
\frac{d\phi}{dr}=\frac{1}{r^{2}}\left[\frac{1}{b^{2}}-\frac{f^{*}}{r^{2}}\right]^{-1/2}.
\]
Knowing the impact parameter $b$ and the function $\frac{d\phi}{dr}$ the we can in principle obtain completely $g^{2}$.

A nonzero value of $g^{2}$ can be also observed measuring the total bending angle, although it will not be possible to reconstruct the whole function. The deflection angle is given by
\[
\delta\phi=2\int_{R_{0}}^{\infty}\frac{dr}{r^{2}}\left[\frac{1}{b^{2}}-\frac{f^{*}}{r^{2}}\right]^{-1/2}-\pi.
\]
Therefore, considering the functions $g_{1}^{2}$ and $g_{2}^{2}$, we have that
\[
\delta\phi_{2}\geq\delta\phi_{1}
\]
is true if the relative radial functions $f_{1}^{*}$ and $f_{2}^{*}$ satisfy the following requirements:
\begin{eqnarray}
f_{1}^{*}(r)\geq f_{2}^{*}(r) &\qquad& \textrm{for }r\geq R_{0} \qquad\textrm{and}\label{f1>f2}\\
\frac{f_{1}^{*}(r)}{r^{2}},\frac{f_{2}^{*}(r)}{r^{2}} &&\textrm{are monotonically decreasing.}\label{f/r2decreasing}
\end{eqnarray}
The above requirements are not the most general but are enough easy to deduce what happens for some of the metrics considered here. First of all we know that if we choose the metric (\ref{fmin}), that is $f_{2}^{*}=f_{min}^{*}=1-\frac{2M}{r}$, the condition (\ref{f1>f2}) is verified for any other possible $f^{*}$. The condition (\ref{f/r2decreasing}) is also verified for $f^{*}_{min}$. So we can conclude that for every $g^{2}$ such that the relative $f^{*}$ satisfies condition (\ref{f/r2decreasing}) the deflection angle is smaller than what we expect from GR.

The case $f^{*}=1$ with $g^{2}=0$ and $M=0$ corresponds to a flat Minkowski space-time giving rise to a zero deflection angle. This means that for an $f^{*}$ satisfying condition (\ref{f/r2decreasing}) and such that $f^{*}>1$ the deflection angle is negative, the force being repulsive.


Let us consider as example the KS metric. Using the expansion for the KS metric for $\omega r^{2}\gg1$, $f^{*}=1-\frac{2M}{r}+\frac{2M^{2}}{\omega r^{4}}$ the deflection angle can be approximated as
\[
\delta\phi
\simeq2\int_{R_{0}}^{\infty}\frac{dr}{r^{2}}\left[
\frac{1}{b_{min}^{2}}-\frac{f^{*}_{min}}{r^{2}}
+\frac{2M^{2}}{\omega}\left(
\frac{1}{R_{0}^{6}}-\frac{1}{r^{6}}
\right)
\right]^{-1/2}-\pi
\]\[
\simeq2\int_{R_{0}}^{\infty}\frac{dr}{r^{2}}\left[
\frac{1}{b_{min}^{2}}-\frac{f^{*}_{min}}{r^{2}}
\right]^{-1/2}\left[
1-\frac{M^{2}}{\omega r^{2}}
\frac{1/R_{0}^{6}-1/r^{6}}{\frac{1}{b_{min}^{2}}-\frac{f^{*}_{min}}{r^{2}}}
\right]-\pi
\]\[
=\delta\phi_{GR}-2\int_{R_{0}}^{\infty}\frac{dr}{r^{2}}\frac{M^{2}}{\omega r^{2}}
\frac{1/R_{0}^{6}-1/r^{6}}
{\left[
\frac{1}{b_{min}^{2}}-\frac{f^{*}_{min}}{r^{2}}
\right]^{3/2}}.
\]

\section{The Singularity}\label{sec:singularity}

In this section we will analyze the behavior of particles near the singular points for a black hole in an asymptotically flat space-time.

The eventual external horizon satisfies the condition \cite{CapassoPolychronakos:GSSSSHG} $f^{*}(r_{h})=0$, so we need first of all to check if there are solutions to the condition $f^{*}>0$. Such a case satisfies conditions (\ref{Fconditions}), therefore we can just analyze the diagonal metric to study its properties. Using relation (\ref{rh<=2M}) we can just restrict to the case $r\leq 2M$, $f^{*}$ being positive for $r>2M$.

At $r=0$ we have that
\[
f^{*}(0)=1-g^{2}(0)
\]
where we used condition (\ref{g2r2}). Moreover the condition $f^{*}>0$ yields
\[
1+\omega r^{2}-g^{2}
>\sqrt{\omega^{2}r^{4}+4\omega Mr-2\omega g^{2}r^{2}}.
\]

Let us consider a $g^{2}$ such that $1+\omega r^{2}>g^{2}$ (otherwise there is a range for which $f^{*}<0$ and hence we have a horizon), then
\[
g^{4}-2g^{2}+1+2\omega r^{2}-4\omega Mr>0
\]
which implies
\[
g^{2}<1-\sqrt{2\omega r^{2}\left(\frac{2M}{r}-1\right)} \qquad
g^{2}>1+\sqrt{2\omega r^{2}\left(\frac{2M}{r}-1\right)}.
\]
The first condition corresponds to a positive $g^{2}$ for $r<M-\sqrt{M^{2}-1/2\omega}$ and $r>M+\sqrt{M^{2}-1/2\omega}$ if $M>\sqrt{\frac{1}{2\omega}}$, while for $M\leq\sqrt{\frac{1}{2\omega}}$ any $g^{2}$ such that $0<g^{2}<1-\sqrt{2\omega r^{2}\left(\frac{2M}{r}-1\right)}$ satisfies all the requirements. The second condition, instead, does not satisfy the condition
\[
1+\omega r^{2}>1+\sqrt{2\omega r^{2}\left(\frac{2M}{r}-1\right)} \quad\Rightarrow\quad
r(\omega^{2}r^{3}+2\omega r-4\omega M)>0
\]
for any $r<2M$ so must be excluded if we want $g^{2}$ to be a continuous function (only in the case $M=0$\footnote{For $M=0$, whatever is the source for $N_{r}$, there are $g^{2}$ for which $f^{*}>0$ but the condition $1+\omega r^{2}$ does not allow to have singularities, being $g^{2}(0)<1$.} we can consider such a case).

Therefore we cannot have vacuum solutions with no horizon other than in the case (we are considering only the expression of $g^{2}$ for $r\leq2M$ because for $r>2M$ we can consider any analytic continuation)
\[
g^{2}<1-\sqrt{2\omega r\left(2M-r\right)} \qquad\textrm{with}\qquad
M<\sqrt{\frac{1}{2\omega}};
\]
note that in this case there is no singularity at $r=0$, $f^{*}(0)=1-g^{2}(0)>0$, only a possible pinch. In particular this means that there are no naked singularities, if we exclude the pinch.

Unlike GR, in HL gravity the 4D metric $g_{\mu\nu}$ is not physically important: the foliation structure is geometrically and physically relevant. In HL gravity the foliation is determined by a scalar function $\phi$ that, for fixed values of the time coordinates, describe a space-like hypersurface $\Sigma$. We can fix the arbitrariness in the coordinates \cite{GermaniKehagiasSfetsos:RQGLP} choosing a guage in which $t=\phi(\overrightarrow{r})$ obtaining the parametrization usually used in HL gravity. The important geometric quantities, other than $\phi$, are the normalized time-like vector $n^{\alpha}$, orthogonal to $\phi$, and the space-like shift vector $N^{\alpha}$ tangent to $\Sigma$. The time direction is then introduced as $t^{\alpha}=n^{\alpha}+N^{\alpha}$. The lapse function is related to the foliation from the definition
\[
n_{\alpha}=-N\partial_{\alpha}\phi.
\]
It is evident then to have a well behaved foliation we need to have a surface $\Sigma$ with a well defined orthogonal vector $n^{\alpha}$. In the gauge $t=\phi$
\[
n_{\alpha}=(-N,0,0,0)
\]
hence the foliation is well defined if $N\neq0$. In our case $N=\pm\sqrt{f}$, thus a foliation is well defined only if $f>0$. Let us call $r_{f}$ the outer radius satisfying the condition $f(r_{f})=0$. Surprisingly we have
\[
f^{*}(r_{f})=f(r_{f})-g^{2}(r_{f})=-g^{2}(r_{f})\leq0,
\]
that is,
\begin{equation}\label{rf<=rh}
r_{f}\leq r_{h}.
\end{equation}
This means that the foliation may also be well defined also behind the horizon. For the KS metric we obviously are in the condition $g^{2}(r_{f})=0$ that implies $r_{f}=r_{h}$, but for any metric such that $g^{2}(r_{f})\neq0$ it is always possible to define a foliation also behind the horizon. We already discussed in section~\ref{sec:constraints_g2} the implications of condition $f>0$ (\ref{f>0}) and they simply implies that if an $r_{f}$ exists is between to horizons, being $f^{*}(r_{f})\leq0$.

As example we have that the metric relative to $f^{*}_{min}$ has a well defined foliation, being $N^{2}=f=1$.

In general it is not clear what happens for $r\leq r_{f}$ because the foliation structure breaks down, introducing a different kind of singularity. To explore what happens to a particle travelling toward $r_{h}$ and then toward $r_{f}$ let us consider a photon of energy $E$ fallowing a radial trajectory. Unlike GR there are no constraints from the fact that a particle is space-like, null-like or time-like because in this context the $4D$-metric has no direct physical meaning (there is not a clear causality structure). Here we will base our discussion on the geometrical properties of space-time in terms of its physical foliation and we will consider the equations of motion for a particle to be the same everywhere, inside or outside the horizons:
\begin{eqnarray}
\textrm{ingoing particle} &\dot{r}=-E, \; \dot{t}^{*}=\frac{E}{N^{*2}}
& \Rightarrow\quad
\dot{t}=-\left[\frac{1}{N^{*2}}-\frac{N_{r}}{f^{*}}\right]\dot{r}\\
\textrm{outgoing particle} &\dot{r}=+E, \;\dot{t}^{*}=\frac{E}{N^{*2}}
& \Rightarrow\quad
\dot{t}=+\left[\frac{1}{N^{*2}}+\frac{N_{r}}{f^{*}}\right]\dot{r}.
\end{eqnarray}

Because we are considering asymptotically flat spherically symmetric space-times, $\lim_{r\to\infty}g^{2}=0$ and hence $\lim_{r\to\infty}f^{*}=\lim_{r\to\infty}f=1$. This implies that outside the outer horizon $f^{*}$ and $f$ are both positive and that the outer horizon $r_{h}^{(0)}$ is the first zero of $f^{*}$. The consequence of this statement is that outside the black hole $N_{r}^{2}<1$ while in general we have
\[
\begin{array}{lcr}
f^{*}>0 & \Rightarrow & N_{r}^{2}<1\\
f^{*}<0 & \Rightarrow & N_{r}^{2}>1.
\end{array}
\]
We will assume that $f^{*'}(r_{h}^{(i)})\neq0$ and $f'(r_{f})\neq0$.

On the horizon, if $r_{f}\neq r_{h}^{(i)}$, $N_{r}^{2}(r_{h}^{(i)})=1$ and we can approximate $N_{r}$ near $r_{h}^{(i)}$ as follows
\[
\begin{array}{l@{\qquad}l}
N_{r}>0: & N_{r}\simeq+1-\frac{1}{2}\frac{f^{*'}(r_{h}^{(i)})}{f(r_{h}^{(i)})}(r-r_{h}^{(i)})\\
N_{r}<0: & N_{r}\simeq-1+\frac{1}{2}\frac{f^{*'}(r_{h}^{(i)})}{f(r_{h}^{(i)})}(r-r_{h}^{(i)}).
\end{array}
\]
If instead $r_{f}=r_{h}^{(i)}$ for a given $i$ then $g^{2}$ near $r_{h}^{(i)}$ goes like $g^{2}\simeq\frac{D^{2n}(g^{2})(r_{h}^{(i)})}{(2n)!}(r-r_{h}^{(i)})^{2n}$, where $[D^{2n}g^{2}](r_{h}^{(i)})$ is the first -~even, being $g^{2}>0$,~- non zero derivative of $g^{2}$ in $r_{h}^{(i)}$. Therefore, near $r_{h}^{(i)}$, $N_{r}$ goes like
\[
N_{r}\simeq\pm\sqrt{\frac{[D^{2n}g^{2}](r_{h}^{(i)})}{f'(r_{h}^{(i)})}\frac{(r-r_{h}^{(i)})^{2n-1}}{(2n)!}};
\]
in particular we have $N_{r}(r_{h}^{(i)})=0$. In the above relation we used the fact that $f^{*'}(r_{h}^{(i)})\neq0$, which implies $f'(r_{h}^{(i)})\neq0$, being $g^{2'}(r_{h}^{(i)})=0$. The case $g^{2}(r)=0$ is then included in the case $g^{2}(r_{h}^{(i)})=0$ considering that all the following derivatives of $g^{2}$ are all zero.

The last case to consider is what happens in $r_{f}$ for $r_{f}\neq r_{h}^{(i)}$ for any $i$. Assuming that $f'(r_{f})\neq0$ we have that $f'(r_{f})>0$ because of the asymptotic flatness. Moreover $f^{*}(r_{f})=-g^{2}(r_{f})<0$, otherwise we fall in the above case for $g^{2}(r_{h}^{(i)})=0$. Then $N_{r}$ near $r_{f}$ is given by
\[
N_{r}\simeq\pm\sqrt{\frac{g^{2}(r_{f})}{f'(r_{f})(r-r_{f})}}
\]
and it is singular in $r_{f}$. This behavior is expected considering that in $r_{f}$ the time direction become tangential and that $N^{-1}$ is singular in $r_{f}$.

Let's start considering a photon near\footnote{We shall consider only the time intervals around the points of interest because we want to show only if they are finite or no, positive or no.} the outer horizon $r_{h}^{(0)}$, for which $f^{*'}(r_{h}^{(0)})>0$, and crossing it from outside:
\[
\begin{array}{l@{\qquad}l}
N_{r}(r_{h}^{(0)})>0: \qquad \Delta t
\simeq-\int_{r_{h}+\delta}^{r}\frac{dr}{2f(r_{h}^{(0)})}
=\frac{r_{h}^{(0)}+\delta-r}{2f(r_{h}^{(0)})}
\\
N_{r}(r_{h}^{(0)})<0: \qquad \Delta t
\simeq-\int_{r_{h}+\delta}^{r}\frac{2dr}{f^{*'}(r_{h}^{(0)})(r-r_{h}^{(0)})}
=\frac{2}{f^{*'}(r_{h}^{(0)})}\ln\left|\frac{\delta}{r-r_{h}^{(0)}}\right|
\\
N_{r}(r_{h}^{(0)})=0: \qquad \Delta t
\simeq-\int_{r_{h}+\delta}^{r}\frac{dr}{f^{*'}(r_{h}^{(0)})(r-r_{h}^{(0)})}
=\frac{1}{f^{*'}(r_{h}^{(0)})}\ln\left|\frac{\delta}{r-r_{h}^{(0)}}\right|
\end{array}
\]
For $N_{r}(r_{h}^{(0)})>0$ we can extend the integral to $r\leq r_{h}^{(0)}$ (in this case $f^{*}<0$) obtaining a finite positive value ($f(r_{h}^{(0)})>0$ by construction), obviously inside the limits for which our approximation is still valid. In the remaining two cases the coordinate time interval goes to $+\infty$ for $r\to r_{h}^{(0)}$. This means that for $N_{r}(r_{h}^{(0)})\leq0$ the black hole behaves just like a Schwarzschild black hole, while for $N_{r}(r_{h}^{(0)})>0$ a particle can cross the horizon in a finite coordinate time and if we consider the limit of integrations to be from $r<r_{h}^{(0)}$ to $r_{h}^{(0)}-\delta$ (in this case $f^{*}<0$) the interval of time becomes negative and is divergent for $r\to r_{h}^{(0)}$. This last statement can be physically interpreted saying that for $N_{r}(r_{h}^{(0)})\leq 0$ particles behind the horizon ($r<r_{h}^{(0)}$) can travel only outward. Will see that this is indeed possible once we shall look to the motion of outgoing particles.

There may exists an other horizon $r_{h}^{(1)}$ just behind $r_{h}^{(0)}$ but, in this case, $f^{*'}(r_{h}^{(1)})<0$. In a similar way we can show that we obtain the same results as before.

If there exist other horizons then we go back considering one of the to above cases.

Following the same steps we find that the situation for an outgoing particle for $f^{*'}(r_{h}^{(0)})>0$ is reversed ($N_{r}>0\Leftrightarrow N_{r}<0$), giving
\[
\begin{array}{l@{\qquad}l}
N_{r}(r_{h}^{(0)})>0: \qquad \Delta t
\simeq\int^{r_{h}+\delta}_{r}\frac{2dr}{f^{*'}(r_{h}^{(0)})(r-r_{h}^{(0)})}
=\frac{2}{f^{*'}(r_{h}^{(0)})}\ln\left|\frac{\delta}{r-r_{h}^{(0)}}\right|
\\
N_{r}(r_{h}^{(0)})<0: \qquad \Delta t
\simeq\int^{r_{h}+\delta}_{r}\frac{dr}{2f(r_{h}^{(0)})}
=\frac{r_{h}^{(0)}+\delta-r}{2f(r_{h}^{(0)})}
\\
N_{r}(r_{h}^{(0)})=0: \qquad \Delta t
\simeq\int^{r_{h}+\delta}_{r}\frac{dr}{f^{*'}(r_{h}^{(0)})(r-r_{h}^{(0)})}
=\frac{1}{f^{*'}(r_{h}^{(0)})}\ln\left|\frac{\delta}{r-r_{h}^{(0)}}\right|
\end{array}
\]
Therefore for $N_{r}(r_{h}^{(0)})\geq0$ the coordinate becomes infinite for $r\to r_{h}^{(0)}$ and in particular it is negative if the particle travels toward $r_{h}^{(0)}$ from inside. As before the case $f^{*'}<0$ gives similar results.

Therefore we can deduce that while $N_{r}(r_{h}^{(i)})>0$ a photon can travel toward the center of the black hole in a finite coordinate time while if $N_{r}(r_{h}^{(i)})<0$ and the photon is an a region in which $f^{*}(r)>0$, the photon take an infinite coordinate time to reach the horizon toward which is traveling. Moreover if $N_{r}(r_{h}^{(i)})<0$ and the photon is a region in which $f^{*}<0$ the photon can travel only toward outside. If the photon is traveling toward outside the situation is completely reversed: particles can come out in a finite coordinate time for $N_{r}(r_{h}^{(i)})<0$ and need an infinite coordinate to move away from an horizon if $f^{*}(r_{h}^{(i)})>0$ and $N_{r}(r_{h}^{(i)})>0$.

If there is an $r_{h}^{(i')}=r_{f}$ then $N_{r}(r_{h}^{(i')})=0$ and this horizon behaves just like in GR, that is, no photon can go away from the horizon surface in a finite coordinate time and no photon reaches the horizon in a finite coordinate time. In this last case the foliation structure also breaks down in this point so we will not worry about what happens inside the horizon. In general we should consider an extension like the Kruskal extension in GR but it is not clear if this procedure is compatible with the theory, corresponding to a non admissible change of coordinates.

The last case to consider is when a photon travels toward $r_{f}\neq r_{h}^{(i)}$. We already pointed out that $f^{*}(r_{f})\leq 0$ (we already studied the case in which the equality is true so will exclude it from the following analysis) therefore we already know that for $N_{r}>0$ we need to consider only photons moving toward $r_{f}$ and for $N_{r}<0$ only photons moving away from $r_{f}$:
\[
\begin{array}{l@{\qquad}l}
N_{r}(r_{f})>0: \qquad \Delta t
\simeq-\int_{r_{f}+\delta}^{r}\frac{dr}{f^{*}(r_{f})}\left[
1-\sqrt{\frac{g^{2}(r_{f})}{f'(r_{f})(r-r_{f})}}
\right]
{\longrightarrow\hspace{-23pt}\raisebox{10pt}{\tiny$r\to r_{f}$}}\frac{1}{f^{*}(r_{f})}\left[
\delta-2\sqrt{\frac{g^{2}(r_{f})\delta}{f'(r_{f})}}
\right]
\\
N_{r}(r_{f})<0: \qquad \Delta t
\simeq\int^{r_{f}+\delta}_{r}\frac{dr}{f^{*}(r_{f})}\left[
1-\sqrt{\frac{g^{2}(r_{f})}{f'(r_{f})(r-r_{f})}}
\right]
{\longrightarrow\hspace{-23pt}\raisebox{10pt}{\tiny$r\to r_{f}$}}\frac{1}{f^{*}(r_{f})}\left[
\delta-2\sqrt{\frac{g^{2}(r_{f})\delta}{f'(r_{f})}}
\right]
\end{array}
\]
The above results are both finite and positive in our approximation ($\delta<4\frac{g^{2}(r_{f})}{f'(r_{f})}$). This means that for $N_{r}(r_{f})>0$ the photon will hit in a finite coordinate time the singularity $r_{f}$ while for $N_{r}(r_{f})<0$ photons can come out of the singularity in a finite coordinate time.

Again behind $r_{f}$ it is not clear if it is possible to extend space-time.

Going back to the case in which $r_{f}=r_{h}^{(i)}$ for a given $i$, like in GR, we obtain that it is necessary a finite proper time to reach the horizon. The KS metric is an example:

\noindent the contribution to the proper time around ($\delta\ll r_{h}^{(0)}$) at the turning point $r_{h}^{(0)}$ for a radially falling (time-like) particle with energy $\varepsilon m$
\[
\Delta\tau=-\int_{r_{h}^{(0)}+\delta}^{r_{h}^{(0)}}\frac{dr}{\sqrt{\varepsilon^{2}-f^{*}}}
\simeq-\int_{r_{h}^{(0)}+\delta}^{r_{h}^{(0)}}\frac{dr}{\sqrt{f^{*'}(r_{h}^{(0)})(r_{h}^{(0)}+\delta-r)}}
=2\sqrt{\frac{\delta}{f^{*'}(r_{h}^{(0)})}}
\]
is finite, being $f^{*'}(r_{h}^{(0)})>0$. In general if $f^{*'}(r_{h}^{(0)})=0$, then the integral is divergent. In particular for an energy $1-(2\omega M^{2})^{1/3}\leq\epsilon<1$, between the two horizons, the motion is periodic with a finite proper time period.

Being the proper time finite we can imagine that something like a Kruskal extension is possible. In GR the Kruscal extension shows that $r_{h}$ is not a singular point but the procedure works because of the general covariance that allows us to consider the same solution in a non-singular coordinate frame system. Here we cannot perform any change of coordinates mixing space and time, so a Kruskal-like extension does not exist. On the contrary it is still possible that a particular interaction term for matter allows only well defined foliations.

%

An other point to consider in introducing an extension is the behavior of the singularity in $r=0$.

Supposing that we are in the conditions for which a particle will hit the center of the system, what happens after the particle hits $r=0$ is unclear because for $M\neq$ the slope of the KS metric goes like
\[
f_{0}^{*'}(r)=2\omega\left(
r-\frac{\omega r+M}{\sqrt{\omega^{2}r^{4}+4\omega Mr}}
\right) \qquad\Rightarrow\qquad
\lim_{r\to0}f_{0}^{*'}(r)=-\infty
\]
showing the presence of a singularity, a pinch (the Ricci scalar near $r=0$ goes like $\mathcal{R}\simeq-\frac{6\sqrt{\omega M}}{r^{3/2}}$).

To have a smooth behavior at $r=0$, that is to have a space-time that looks locally flat at $r=0$ letting the particle to go through, we need $f^{*}(0)$ to be finite and $f^{*'}(0)=0$. The first condition implies that $g^{2}(0)$ is finite while the second reduces to
\[
\left[
2\omega r
-\frac{2\omega^{2}r^{3}+2\omega M-2\omega g^{2}r-\omega(g^{2})'r^{2}}
{\sqrt{\omega^{2}r^{4}+4\omega Mr-2\omega g^{2}r^{2}}}
-(g^{2})'
\right]_{r=0}=0.
\]
For $r\simeq0$, $f^{*'}$ reduces to
\[
2\omega r
-\frac{2\omega M-\omega(g^{2})'r^{2}}
{\sqrt{4\omega Mr}}
-(g^{2})'\simeq0,
\]
that is,
\[
(g^{2})'\simeq-\sqrt{\frac{\omega M}{r}} \qquad\Rightarrow\qquad
g^{2}\simeq-2\sqrt{\omega Mr}
\]
showing that we cannot have a smooth behavior at the origin for $M\neq0$, then the presence of a point-mass still correspond to a singularity in space-time.

If we consider the case $M=0$ with $N_{r}\neq0$, then
\[
\left[
2\omega r
-\frac{2\omega^{2}r^{3}-2\omega g^{2}r-\omega(g^{2})'r^{2}}
{\sqrt{\omega^{2}r^{4}-2\omega g^{2}r^{2}}}
-(g^{2})'
\right]_{r=0}=0.
\]
For $r\simeq0$, $f^{*'}$ reduces to ($g^{2}<\omega r^{2}/2$ for $r\simeq0$)
\[
2\omega r
-\frac{2\omega^{2}r^{3}-2\omega g^{2}r-\omega(g^{2})'r^{2}}
{\omega r^{2}}
-(g^{2})'\simeq0,
\]
that is, $\left.\frac{g^{2}}{r}\right|_{r=0}=0$. This property means that are possible locally non-flat vacuum solutions with $M=0$ and $N_{r}\neq0$ and smooth in $r=0$. In this case there must be some other source, other than $M$ responsible for an $N_{r}\neq0$. This possibility will then depend strictly on the particular coupling with matter.

\section{Conclusions}

The reduced symmetries of HL gravity make unclear the meaning of the symmetry (\ref{deltag}) from a physical point of view. Such a symmetry may be just an accident in the current formulation of the theory, but it can be used, if generalized, to fix $\lambda$ to the value of $1$ in the quantization process.

In the meanwhile it is evident that for standard relativistic matter the symmetry (\ref{deltag}) is not a symmetry. Although in principle we can construct an interaction term that is invariant under such a gauge symmetry, the physical consequence is that for a relativistic coupling with matter the symmetry is broken: therefore every value of the function $g^{2}$ corresponds to a a different solution.

In this paper we studied the constraints that $g^{2}$ must satisfies in order to have a well defined metric and to satisfies physical requests and we have analyzed what are the implications on the trajectories of particles (in some cases only massless).

Not having the full relativistic symmetry we use as starting point the dynamic of particles and analyze the behavior of their trajectories. We consider a possible way to reconstruct $g^{2}$ measuring the bending of light to reconstruct the metric. Moreover we analyze the motion of massless particle in the presence of a black hole.

Here we do not consider any model for the collapse so we do not worry if it is possible to have trapped particle between two horizons during the collapse but we simply analyze how long it takes to move toward to or away from a horizon. As simple consequence we have that if a black hole has a radial shift vector toward outside then massless particles can travel in a finite coordinate time toward inside, while if the shift vector is directed inwardly massless particles can came out in a finite coordinate time.

Finally we suggest a possible redefinition of singularity. The time-direction is not well defined for $N=0$ then we can identify such a point as a break down of the foliation. In the spherically symmetric case such a point is defined by the condition $f(r_{f})=0$ (we need to consider only the outer radius satisfying this property). For the KS metric such a point corresponds with the outer horizon. With this definition we do not need to ask for any extension of the metric behind such a point because the geometric structure, the foliation, is not well defined, moreover, unlike GR,we are not supposed to consider the problem in a different set of coordinates because it would be unphysical. It is still possible that a particular coupling with matter or perhaps also the standard one would imply, once the collapse is studied, that the foliation is always well defined under certain physical conditions.

\appendix

\section{Appendix}\label{sec:appendix}

As pointed out in \cite{CapassoPolychronakos:GSSSSHG} the most generic spherically symmetric metric is with a nonzero shift variables and is given by (\ref{ansatz})
\[
ds^{2}=
-(N^{2}-N_{r}^{2}f)dt^{2}
+2N_{r}drdt
+\frac{dr^{2}}{f}
+r^{2}d\theta^{2}+r^{2}\sin^{2}\theta d\phi^{2}.
\]
In General Relativity we can always perform the following change of coordinates
\begin{equation}\label{spacetimereparametrization}
dt=dt^{*}+F(r)dr^{*} \qquad r=r^{*}
\end{equation}
obtaining
\[
ds^{2}=
-(N^{2}-N_{r}^{2}f)dt^{*2}
+2[N_{r}-(N^{2}-N_{r}^{2}f)F]dr^{*}dt^{*}+
\]\[
+\left[
\frac{1}{f}-(N^{2}-N_{r}^{2}f)F^{2}+2N_{r}F
\right]dr^{*2}
+r^{*2}d\theta^{2}+r^{*2}\sin^{2}\theta d\phi^{2}.
\]
Choosing
\begin{equation}\label{Fgeneral}
F=\frac{N_{r}}{N^{2}-N_{r}^{2}f}
\end{equation}
and defining
\[
N^{*2}=N^{2}-N_{r}^{2}f \qquad
f^{*}=\frac{f(N^{2}-N_{r}^{2}f)}{N^{2}}
\]
we have that the metric takes the usual diagonal form (\ref{diagonalmetric}):
\[
ds^{2}=
-N^{*2}dt^{*2}
+\frac{1}{f^{*}}dr^{*2}
+r^{*2}d\theta^{2}+r^{*2}\sin^{2}\theta d\phi^{2}.
\]
Moreover note that if $f=N^{2}$ we also have
\[
\frac{f^{*}}{N^{*2}}=\frac{f}{N^{2}}=1.
\]

Unlike GR, in HL gravity we cannot perform the change of coordinates (\ref{spacetimereparametrization}) because such a transformation does not preserve the foliation $M=\mathbb{R}\times\Sigma$ of spacetime. Indeed, because of the anisotropy, the theory is invariant only under diffeomorphisms that leave unchanged the foliation structure (\cite{Lawson:F,MoerdijkMrcun:IFLG}) $\mathcal{F}$:
\[
x^{i}\to \tilde{x}^{i}=\tilde{x}^{i}(x,t) \qquad
t\to \tilde{t}=\tilde{t}(t).
\]


\vskip 0.2in
\noindent
\emph{\underline {Acknowledgements}:} 
I wish to thank my advisor A.~P.~Polychronakos for all the comments and useful discussions.


\begin{thebibliography}{99}
\bibitem{Horava:MQC}
P.~Horava,
``Membranes at Quantum Criticality,''
JHEP {\bf 0903}, 020 (2009)
\href{http://arxiv.org/abs/0812.4287}{[arXiv:0812.4287 [hep-th]]}.

\bibitem{Horava:QGLP}
P.~Horava,
``Quantum Gravity at a Lifshitz Point,''
Phys.\ Rev.\  D {\bf 79}, 084008 (2009)
\href{http://arxiv.org/abs/0901.3775}{[arXiv:0901.3775 [hep-th]]}.

\bibitem{KiritsisKofinas:HLC}
E.~Kiritsis and G.~Kofinas,
``Horava-Lifshitz Cosmology,''
Nucl.\ Phys.\  B {\bf 821}, 467 (2009)
\href{http://arxiv.org/abs/0904.1334}{[arXiv:0904.1334 [hep-th]]}.

\bibitem{SotiriouVisserWeinfurtner:PLVQG}
T.~P.~Sotiriou, M.~Visser and S.~Weinfurtner,
``Phenomenologically viable Lorentz-violating quantum gravity,''
Phys.\ Rev.\ Lett.\  {\bf 102}, 251601 (2009)
\href{http://arxiv.org/abs/0904.4464}{arXiv:0904.4464 [hep-th]}.

\bibitem{SotiriouVisserWeinfurtner:QGWLI}
T.~P.~Sotiriou, M.~Visser and S.~Weinfurtner,
``Quantum gravity without Lorentz invariance,''
JHEP {\bf 0910}, 033 (2009)
\href{http://arxiv.org/abs/0905.2798}{arXiv:0905.2798 [hep-th]}

\bibitem{KehagiasSfetsos:BHFRWGNRG}
A.~Kehagias and K.~Sfetsos,
``The black hole and FRW geometries of non-relativistic gravity,''
Phys.\ Lett.\  B {\bf 678}, 123 (2009)
\href{http://arxiv.org/abs/0905.0477}{[arXiv:0905.0477 [hep-th]]}.

\bibitem{CapassoPolychronakos:GSSSSHG}
D.~Capasso and A.~P.~Polychronakos,
``General static spherically symmetric solutions in Horava gravity,''
Phys.\ Rev.\  D {\bf 81}, 084009 (2010)
\href{http://arxiv.org/abs/0911.1535}{[arXiv:0911.1535 [hep-th]]}.

\bibitem{CharmousisNizPadillaSaffin:SCHG}
C.~Charmousis, G.~Niz, A.~Padilla and P.~M.~Saffin,
``Strong coupling in Horava gravity,''
JHEP {\bf 0908}, 070 (2009)
\href{http://arxiv.org/abs/0905.2579}{[arXiv:0905.2579 [hep-th]]}.

\bibitem{BlasPujolasSibiryakov:EMIHG}
D.~Blas, O.~Pujolas and S.~Sibiryakov,
``On the Extra Mode and Inconsistency of Horava Gravity,''
\href{http://arxiv.org/abs/0906.3046}{arXiv:0906.3046 [hep-th]}.

\bibitem{Park:BHCSIRMHG}
M.~i.~Park,
``The Black Hole and Cosmological Solutions in IR modified Horava Gravity,''
JHEP {\bf 0909}, 123 (2009)
\href{http://arxiv.org/abs/0905.4480}{[arXiv:0905.4480 [hep-th]]}.

\bibitem{Park:BHSIRMHG}
M.~i.~Park,
``The Black Hole and Cosmological Solutions in IR modified Horava Gravity,''
\href{http://arxiv.org/abs/0905.4480}{arXiv:0905.4480 [hep-th]}.

\bibitem{CaiSaridakis:NSCNRG}
Y.~F.~Cai and E.~N.~Saridakis,
``Non-singular cosmology in a model of non-relativistic gravity,''
JCAP {\bf 0910}, 020 (2009)
\href{http://arxiv.org/abs/0906.1789}{[arXiv:0906.1789 [hep-th]]}.

\bibitem{AliDuttaSaridakisSen:HLCCG}
A.~Ali, S.~Dutta, E.~N.~Saridakis and A.~A.~Sen,
``Horava-Lifshitz cosmology with generalized Chaplygin gas,''
\href{http://arxiv.org/abs/1004.2474}{arXiv:1004.2474} [astro-ph.CO].

\bibitem{Konoplya:TCHLG}
R.~A.~Konoplya,
``Towards constraining of the Horava-Lifshitz gravities,''
Phys.\ Lett.\  B {\bf 679}, 499 (2009)
\href{http://arxiv.org/abs/0905.1523}{[arXiv:0905.1523 [hep-th]]}.

\bibitem{HarkoKivacsLobo:SSTHLG}
T.~Harko, Z.~Kovacs and F.~S.~N.~Lobo,
``Solar system tests of Ho\v{r}ava-Lifshitz gravity,''
\href{http://arxiv.org/abs/0908.2874}{arXiv:0908.2874 [gr-qc]}.

\bibitem{Park:THGDE}
M.~i.~Park,
``A Test of Horava Gravity: The Dark Energy,''
\href{http://arxiv.org/abs/0906.4275}{arXiv:0906.4275 [hep-th]}.

\bibitem{IorioRuggiero:HLGSSOM}
L.~Iorio and M.~L.~Ruggiero,
``Phenomenological constraints on the Kehagias-Sfetsos solution in the Horava-Lifshitz gravity from solar system orbital motions,''
\href{http://arxiv.org/abs/0909.2562}{arXiv:0909.2562 [gr-qc]}.

\bibitem{IorioRuggiero:CKSHLG}
L.~Iorio and M.~L.~Ruggiero,
``Constraining the Kehagias-Sfetsos solution in the Horava-Lifshitz gravity with extrasolar planets,''
\href{http://arxiv.org/abs/0909.5355}{arXiv:0909.5355 [gr-qc]}.

\bibitem{LiPang:THLG}
M.~Li and Y.~Pang,
``A Trouble with Ho\v{r}ava-Lifshitz Gravity,''
\href{http://arxiv.org/abs/0905.2751}{arXiv:0905.2751 [hep-th]}.

\bibitem{Kobakhidze:IRHGHC}
A.~Kobakhidze,
``On the infrared limit of Horava's gravity with the global Hamiltonian constraint,''
Phys.\ Rev.\  D {\bf 82}, 064011 (2010)
\href{http://arxiv.org/abs/0906.5401}{[arXiv:0906.5401 [hep-th]]}.

\bibitem{ChenHuang:FTLP}
B.~Chen and Q.~G.~Huang,
``Field Theory at a Lifshitz Point,''
\href{http://arxiv.org/abs/0904.4565}{arXiv:0904.4565 [hep-th]}.

\bibitem{OrlandoReffert:RHLG}
D.~Orlando and S.~Reffert,
``On the Renormalizability of Horava-Lifshitz-type Gravities,''
Class.\ Quant.\ Grav.\  {\bf 26}, 155021 (2009)
\href{http://arxiv.org/abs/0905.0301}{[arXiv:0905.0301 [hep-th]]}.

\bibitem{Suyama:2009vy}
T.~Suyama,
``Notes on Matter in Horava-Lifshitz Gravity,''
\href{http://arxiv.org/abs/0909.4833}{arXiv:0909.4833 [hep-th]}.

\bibitem{CapassoPolychronakos}
D.~Capasso and A.~P.~Polychronakos,
``Particle Kinematics in Horava-Lifshitz Gravity,''
\href{http://arxiv.org/abs/0909.5405}{arXiv:0909.5405 [hep-th]}.

\bibitem{Sindoni:PKHL}
L.~Sindoni,
``A note on particle kinematics in Horava-Lifshitz scenarios,''
\href{http://arxiv.org/abs/0910.1329}{arXiv:0910.1329 [gr-qc]}.

\bibitem{Rama:PMHLDR}
S.~K.~Rama,
``Particle Motion with Ho\v{r}ava -- Lifshitz type Dispersion Relations,''
\href{http://arxiv.org/abs/0910.0411}{arXiv:0910.0411 [hep-th]}.

\bibitem{Suyama:NMHLG}
T.~Suyama,
``Notes on Matter in Horava-Lifshitz Gravity,''
\href{http://arxiv.org/abs/0909.4833}{arXiv:0909.4833 [hep-th]}.

\bibitem{RomeroCuestaGarciaVergara:CAM}
J.~M.~Romero, V.~Cuesta, J.~A.~Garcia and J.~D.~Vergara,
``Conformal Anisotropic Mechanics,''
\href{http://arxiv.org/abs/0909.3540}{arXiv:0909.3540 [hep-th]}.

\bibitem{LuMeiPope:SHG}
H.~Lu, J.~Mei and C.~N.~Pope,
``Solutions to Horava Gravity,''
Phys.\ Rev.\ Lett.\  {\bf 103}, 091301 (2009)
\href{http://arxiv.org/abs/0904.1595}{[arXiv:0904.1595 [hep-th]]}.

\bibitem{MyungKim:THLBH}
Y.~S.~Myung and Y.~W.~Kim,
``Thermodynamics of Ho\v{r}ava-Lifshitz black holes,''
\href{http://arxiv.org/abs/0905.0179}{arXiv:0905.0179 [hep-th]}.

\bibitem{CaiCaoOhta:TBHHLG}
R.~G.~Cai, L.~M.~Cao and N.~Ohta,
``Topological Black Holes in Horava-Lifshitz Gravity,''
Phys.\ Rev.\  D {\bf 80}, 024003 (2009)
\href{http://arxiv.org/abs/0904.3670}{[arXiv:0904.3670 [hep-th]]}.

\bibitem{CaiCaoOhta:ThBHHLG}
R.~G.~Cai, L.~M.~Cao and N.~Ohta,
``Thermodynamics of Black Holes in Horava-Lifshitz Gravity,''
Phys.\ Lett.\  B {\bf 679}, 504 (2009)
\href{http://arxiv.org/abs/0905.0751}{[arXiv:0905.0751 [hep-th]]}.

\bibitem{GhodsiHatefi:ERSHG}
A.~Ghodsi and E.~Hatefi,
``Extremal rotating solutions in Horava Gravity,''
\href{http://arxiv.org/abs/0906.1237}{arXiv:0906.1237 [hep-th]}.

\bibitem{Myung:TBHDHLG}
Y.~S.~Myung,
``Thermodynamics of black holes in the deformed Ho\v{r}ava-Lifshitz gravity,''
Phys.\ Lett.\  B {\bf 678}, 127 (2009)
\href{http://arxiv.org/abs/0905.0957}{[arXiv:0905.0957 [hep-th]]}.

\bibitem{Myung:EBHDHLG}
Y.~S.~Myung,
``Entropy of black holes in the deformed Ho\v{r}ava-Lifshitz gravity,''
\href{http://arxiv.org/abs/0908.4132}{arXiv:0908.4132 [hep-th]}.

\bibitem{LeeKimMyung:EBHHLG}
H.~W.~Lee, Y.~W.~Kim and Y.~S.~Myung,
``Extremal black holes in the Ho\v{r}ava-Lifshitz gravity,''
\href{http://arxiv.org/abs/0907.3568}{arXiv:0907.3568 [hep-th]}.

\bibitem{PengWu:HRBHIMHLG}
J.~J.~Peng and S.~Q.~Wu,
``Hawking Radiation of Black Holes in Infrared Modified Ho\v{r}ava-Lifshitz Gravity,''
\href{http://arxiv.org/abs/0906.5121}{arXiv:0906.5121 [hep-th]}.

\bibitem{KiritsisKofinas:HLBH}
E.~Kiritsis and G.~Kofinas,
``On Horava-Lifshitz 'Black Holes',''
\href{http://arxiv.org/abs/0910.5487}{arXiv:0910.5487 [hep-th]}.

\bibitem{TangChen:SSSSMHLGPC}
J.~Z.~Tang and B.~Chen,
``Static Spherically Symmetric Solutions to modified Horava-Lifshitz Gravity with Projectability Condition,''
\href{http://arxiv.org/abs/0909.4127}{arXiv:0909.4127 [hep-th]}.

\bibitem{CaiLiuSun:z=4HLG}
R.~G.~Cai, Y.~Liu and Y.~W.~Sun,
``On the z=4 Horava-Lifshitz Gravity,''
JHEP {\bf 0906}, 010 (2009)
\href{http://arxiv.org/abs/0904.4104}{[arXiv:0904.4104 [hep-th]]}.

\bibitem{Park:HGCP}
M.~i.~Park,
``Horava Gravity and Gravitons at a Conformal Point,''
\href{http://arxiv.org/abs/0910.5117}{arXiv:0910.5117 [hep-th]}.

\bibitem{KoutsoumbasPapantonopoulosPasipoularidesTsoukalas:BHS5DHLG}
G.~Koutsoumbas, E.~Papantonopoulos, P.~Pasipoularides and M.~Tsoukalas,
``Black Hole Solutions in 5D Horava-Lifshitz Gravity,''
Phys.\ Rev.\  D {\bf 81}, 124014 (2010)
\href{http://arxiv.org/abs/1004.2289}{[arXiv:1004.2289 [hep-th]]}.

\bibitem{KoutsoumbasPasipoularides:BHSHLGC}
G.~Koutsoumbas and P.~Pasipoularides,
``Black hole solutions in Horava-Lifshitz Gravity with cubic terms,''
Phys.\ Rev.\  D {\bf 82}, 044046 (2010)
\href{http://arxiv.org/abs/1006.3199}{[arXiv:1006.3199 [hep-th]]}.

\bibitem{HoravaThompson:GCQGLP}
P.~Horava and C.~M.~Melby-Thompson,
``General Covariance in Quantum Gravity at a Lifshitz Point,''
Phys.\ Rev.\  D {\bf 82}, 064027 (2010)
\href{http://arxiv.org/abs/1007.2410}{[arXiv:1007.2410 [hep-th]]}.

\bibitem{AlexandrePasipoularides:SSSCHLG}
J.~Alexandre and P.~Pasipoularides,
``Spherically symmetric solutions in Covariant Horava-Lifshitz Gravity,''
\href{http://arxiv.org/abs/1010.3634}{arXiv:1010.3634 [hep-th]}.

\bibitem{BellorinRestuccia:CHT}
J.~Bellorin and A.~Restuccia,
``On the consistency of the Horava Theory,''
\href{http://arxiv.org/abs/1004.0055}{arXiv:1004.0055 [hep-th]}.

\bibitem{GermaniKehagiasSfetsos:RQGLP}
C.~Germani, A.~Kehagias and K.~Sfetsos,
``Relativistic Quantum Gravity at a Lifshitz Point,''
JHEP {\bf 0909}, 060 (2009)
\href{http://arxiv.org/abs/0906.1201}{[arXiv:0906.1201 [hep-th]]}.

\bibitem{Lawson:F}
H.~B.~Lawson, Jr., ``Foliations'', Bull. Amer. Math. Soc. 80 (1974) 369.

\bibitem{MoerdijkMrcun:IFLG}
I. Moerdijk and J. Mr\v{c}un, ``Introduction to Foliations and Lie Groupoids'', Cambridge U.P. (2003).

\end{thebibliography}
\end{document}